\documentclass[letterpaper]{article}
\usepackage{natbib,alifeconf}

\title{Entropy Walker, a Fast Algorithm for Small Community Detection in Large Graphs}
\author{Luis Argerich$^{1}$ \\
\mbox{}\\
$^1$University of Buenos Aires (U.B.A), CS Department F.I.U.B.A \\
}

\begin{document}
\maketitle

\begin{abstract}
This report presents a very simple algorithm for overlaping community-detection in large graphs under constraints such as the minimum and maximum number of members allowed. The algorithm is based on the simulation of random walks and measures the entropy of each random walk to detect the discovery of a community.
\end{abstract}

\section{Introduction}
Community detection in large graphs is getting attention as an important application of Social Network Analysis (SNA), the ability to detect closely knit communities 
opens several applications from targeting ads to recommender systems. In this work we try to derive a very simple and efficient algorithm for community detection based on a size parameter. Being able to specify the minimum and maximum size of communities to detect can be a critical factor in the SNA area, some networks tend to form very small and dense communities while other networks form larger groups. The first section of this report discusses some existing algorithms for community detection in social graphs, then we introduce the idea behind the entropy walker and present our algorithm. The final sections show some examples of the algorithm being used in some toy examples and analyzes the scaling of the method for large graphs.

\section{Previous Work}
Several algorithms have been developed for community detection in large graphs. Clutsering methods based in k-means need to know in advance the number of communities to find in the network. In practice this is not possible as the number of communities is usually unknown and furthermore due to social interactions the number of communities in a network might change over time making it very hard to set up as a parameter. 

The modularity optimization algorithm [B08] automatically detects the number of communities but it doesn't allow for overlapping communities. This is also inpractical for Social Networks as most nodes will be members of several different social circles.

BigClam [Lesk13] is a fast algorithm to detect overlaping communities, it's based in non-negative matrix factorization but it needs to know the number of communities to detect, as mentioned before this is an important limitation.

[McA13] presents an algorithm to find social circles in networks but is based on node parameters "features", we would like to perform the extraction of communities based in network structure only.

The idea of random walks being used to detect communities is also used in the MCL algorithm [vDon99] however MCL can't control the size of the communities being detected and it needs to perform operations on the complete matrix of the graph limiting its use to small and medium sized networks.

\section{Description}
We define a "tour" as a random walk of length "s". The basic idea of the algorithm is to perform several tours starting from random nodes and to detect communities based on the result of those tours. "s" should be longer than the minimum number of members that we want for a community and it serves as an upper bound for the maximum number of members in a community. 

It is likely for a random walker to get "trapped" inside nodes of a community, going back and forth between them because there are more inter-community edges than edges that will take the walker outside of the community. Even if the random walker goes outside the community chances are it might come back. 

The algorithm will filter the random walks that aren't likely to have found a community calculating the entropy of the tour [Sha48]. Tours with high entropy are unlikely to contain a community because they visit mostly different nodes. They are probably paths or bridges between communities and might be of interest for some other applications.

The entropy is computed using the very popular Shannon formula:

$$H=\sum^n P_i*log(1/P_i)$$

Where $P_i$ is just the probability of the node in the tour, in other words its frequency in the tour over the sum of all node frequencies. A threshold parameter establishes the maximum entropy for a tour to be accepted as a fraction of the maximum possible entropy that can be computed assuming a random walk that never visits the same node more than once. We call this parameter $et$ for entropy threshold.

When $et$ is 1 all the tours are accepted, lowering $et$ increases the amount of rejected tours. The graph in figure 1 shows the percentage of accepted tours for different values of $et$ using the Food Network as an example.

\begin{figure}[!htb]
\begin{center}
\includegraphics[width=3in]{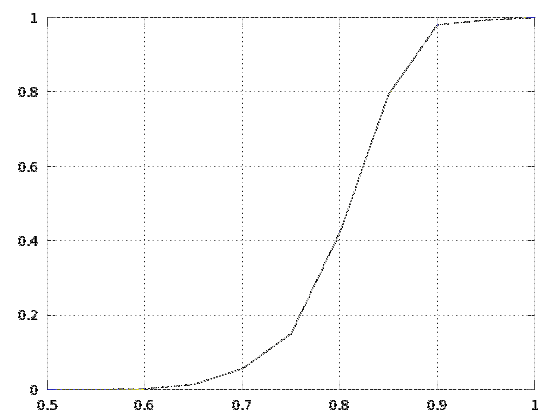}
\caption{Number of tours per entropy threshold.}
\label{fig1}
\end{center}
\end{figure}

The $et$ parameter can be tuned based on two different goals. One possibility is to use it to limit the total number of tours to store in memory for very large graphs, a second use, more logical, is to set how dense a community has to be to be considered. This second use that is data dependant is probably the recommended one.

This is an example of a very low entropy tour from the food network: $ [cream-egg-cream-milk-cream-butter-raisin-vanilla-butter-raisin-cream-butter-cream-vanilla-egg-butter-cream-butter-egg-milk-butter-cream-milk-egg-milk-raisin-milk-vanilla-milk-yogurt]$

And this is an example of a high entropy tour from the same network: $[thyme-tomato-turmeric-carrot-beef-vinegar-beef-garlic-lamb-onion-chicken-ginger-cilantro-coriander-mint-parsley-bread-bell_pepper-cayenne-garlic-lamb-cinnamon-ginger-cumin-ginger-honey-cinnamon-orange_juice-vanilla-raisin]$

We can see how the first tour can be converted in a community with the top ingredients being used for the same kind of dishes, the second tour has a wide array of ingredients and can't be considered a community. Maybe a bridge between different communities. As we have mentioned extracting the high entropy tours from a network might also be an interesting application.

After accepting or rejecting a tour based on its entropy the algorithm will try to see if this tour is new or if it is similar to an already seen tour. Locality sensitive hashing (LSH) can be used to make similar tours hash to the same bucket avoiding the need to compare new tours with the existing ones. If LSH maps the tour to a bucket where a tour is already stored then both tours are merged adding the frequencies of the nodes present in both tours. This greatly reduces the number of tours that need to be stored in memory and avoids the problem of two very similar tours being detected as different communities. 

In some applications the $n$ most frequent nodes in a tour can be used as the key to a hash function to determine the bucket number for the node. This is a simplification of LSH using only one minhash computed from the most frequent nodes in a tour. When this is not possible or doesn't work standard LSH can be used.

Now we describe the parameters used in the algorithm:

The algorithm uses several parameters to fine-tune its behaviour:

\begin{table}[!hbt]
\center{
\begin{tabular}{|c|c|c|}\hline
Parameter & Description \\ \hline\hline
nt & Number of random tours to simulate \\
lt & Length of each simulated tour \\
minm & Minimum number of members for a community \\
maxm  & Maximum number of members for a community \\
et & Entropy threshold for a tour to be a community \\ \hline
\end{tabular}
}
\vskip 0.25cm
\caption{Algorithm parameters.}
\end{table}

The algorithm will perform $nt$ random tours and check the entropy of each tour. If the tour entropy is below the $et$ threshold then the tour will be stored in a hash table along with a counter merging the tour with the already existing one if the bucket is not empty.
It's easy to notice that this process can be parallelized and that several million tours can be performed efficiently. The memory cost to store the tours depends on the algorithm parameters.

When the $et$ (entropy threshold) parameter is low the algorithm with detect only a few very dense communities and tours with frequency 1 can be considered a community. When the $et$ parameter is higher the algorithm will check many tours and it might make sense to discard the tours with lower frequencies keeping the ones that have been repetedly matched. 

\section{A Centrality Measure}

It is known that MonteCarlo Random Walks can be used to compute PageRank and/or Eigenvector centrality, the procedure used to detect communities can be used to compute at the same time a centrality score for the network nodes. So the first conclusion is that node centrality can be computed at the same time as the community detection algorithm runs, just adding 1 to a counter every time a node is visited by a tour and then normalizing the cummulative score.

The effect of entropy filter is show in Fig2. We can see that some nodes produce peaks for entropy thresholds below 1.00, this means that the centrality of those nodes is higher in the entropy filtered sets compared to the plain random walks without filtering. These peaks can be detected computing the delta between the eigenvector centrality and the tour computed centrality. From these peaks we can detect nodes that are both central to the network and to the small communities where they belong, this gives an index of in-community centrality. 

Testing the procedure on the Facebook Ego Network the peaks matched nodes that had a high degree of connections with the members in their communities.

\begin{figure}[!htb]
\begin{center}
\includegraphics[width=3.5in]{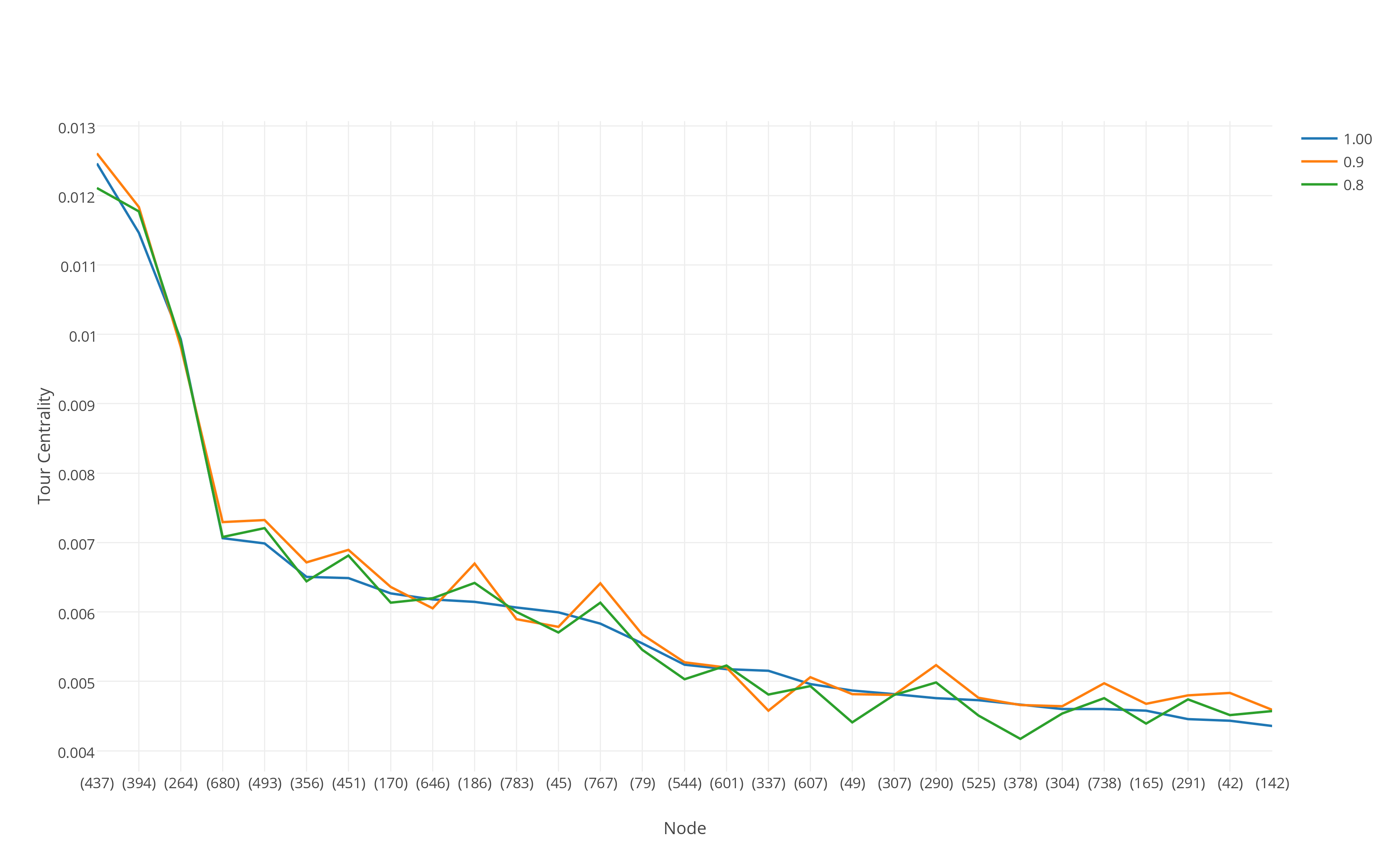}
\caption{Centrality Score for Different Entropy Thresholds.}
\label{fig2}
\end{center}
\end{figure}

\section{Personalized Circles}

Something interesting to notice is that the algorithm can be run starting always from the same node, in the style of a personalized PageRank, when that happens we get as a result the social circles of a given user. This is in some way similar to the algorithm used by Twitter to recommend users to follow[Gup12] the difference is that instead of computing a score for each node we compute scores for each random walk (tour) performed by the simulation. 

For example we can run the algorithm from the Tomato ingredient to see what goes well with Tomato:

\begin{table}[!hbt]
\center{
\begin{tabular}{|l|lp{7cm}}\hline
\parbox{7cm}{(226):[tomato,  garlic, coriander, cayenne, lamb, bread,mint, cucumber]} \\ \hline
\parbox{7cm}{(169):[tomato, bread, cucumber, garlic, parsley, mint, lamb, yogurt]} \\ \hline
\parbox{7cm}{(105):[tomato, bread, lamb, mint, parsley, cucumber, cayenne, carrot]} \\ \hline
\parbox{7cm}{(93):[tomato, parsley, bread, cucumber, mint, cayenne, dill, oliveoil]} \\ \hline
\parbox{7cm}{(65):[tomato, bread, lamb, mint, cucumber, oliveoil, cayenne, vinegar]} \\ \hline
\parbox{7cm}{(55):[tomato, garlic, beef, bread, lamb, cayenne, vinegar, cucumber]} \\ \hline
\parbox{7cm}{(36):[tomato, beef, onion, carrot, lamb, chicken, chickenbroth, bread]} \\ \hline
\end{tabular}
}
\end{table}

Instantaneous delicious recipes!

\section{Analysis}

This section presents some analysis and graphs about the behaviour of the algorithm.

\subsection{Growth of the number of communities for a fixed entropy threshold}

It is interesting to analyze the number of tours that the algorithm will keep in memory as the network grows larger for a constant fixed entropy threshold. We found that the number of tours analyzed does not grow as the size of the network and is strongly dependant on network structure.

\begin{figure}[!htb]
\begin{center}
\includegraphics[width=3.5in]{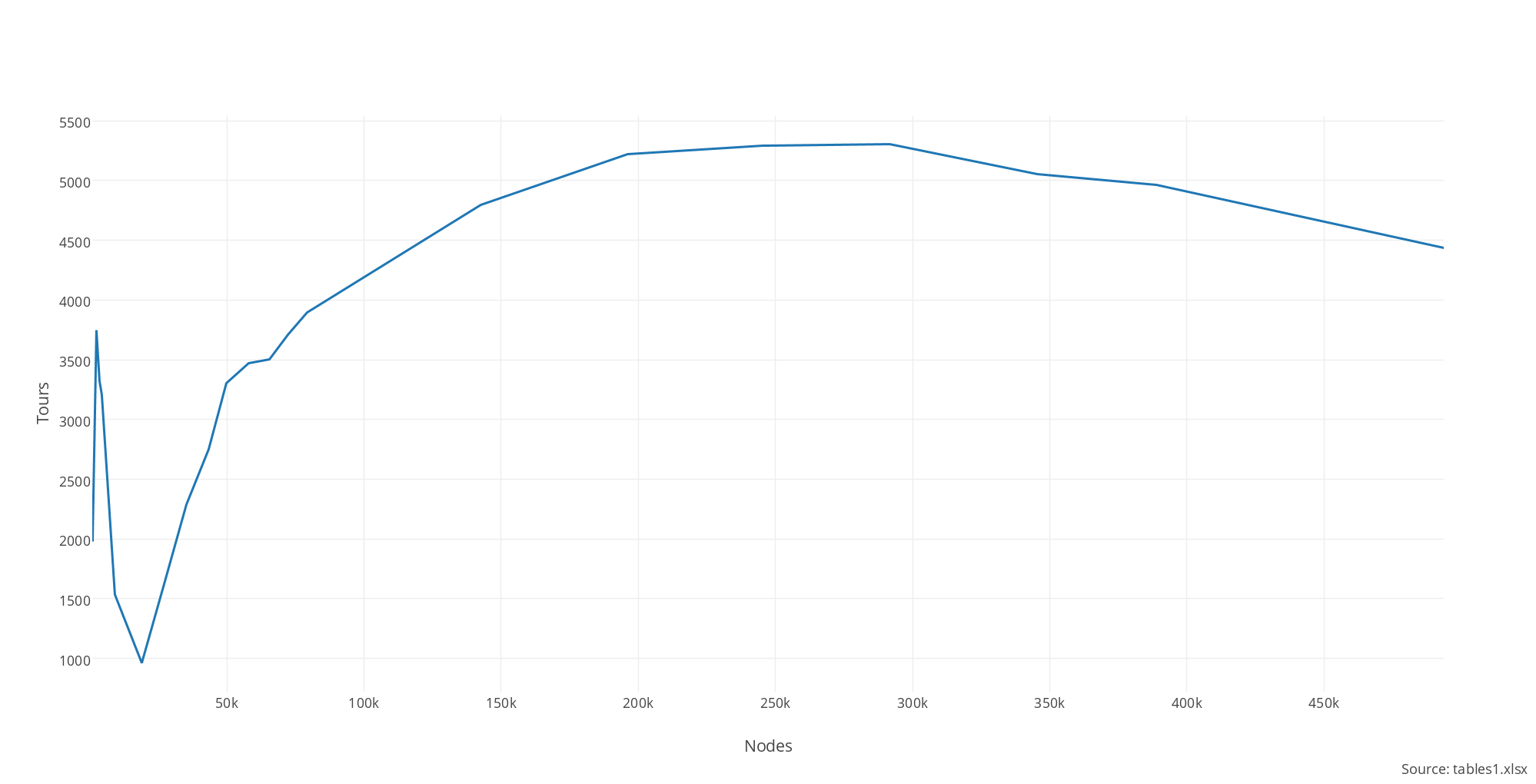}
\caption{Number of Tours per number of Nodes.}
\label{fig3}
\end{center}
\end{figure}

With only a few nodes small communities are common in a graph with high clustering, as the network grows larger the number of small communities quickly goes down. This can be explained because a random walker has now more options and is less likely to get trapped inside a community. Then after more nodes are added a threshold is passed and small communities emerge again. This curious behaviour in the formation of small communities as the network grows larger resulted an interesting find and can be useful to refine generic models for network growth.

\subsection{Relationship to Clustering}

The emergence of small communities in large networks is strongly related to the clustering coefficient of the network. When the clustering coefficient is very los there are not enough edges to form dense communities so small communities will not form in random networks. In the same way if the clustering coefficient is too high then the random walker can visit almost any node from any node and thus will not get trapped inside a small community, the whole network is the only existing community. 

The following graph shows the number of tours detected for a fixed entropy threshold depending on the clustering coefficient of networks synthetically generated using the Barabasi-Albert model[Bar99]. 

\begin{figure}[!htb]
\begin{center}
\includegraphics[width=3.5in]{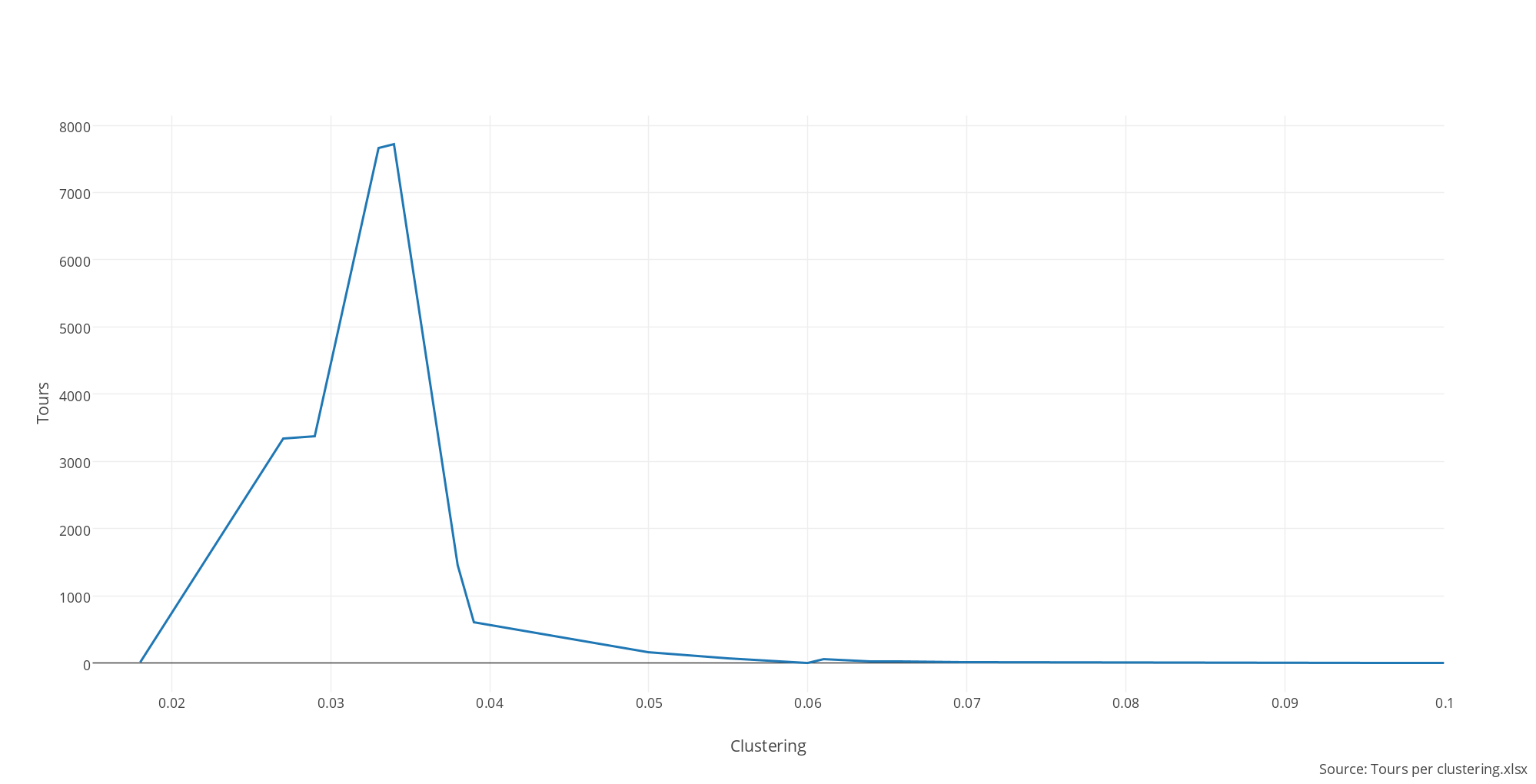}
\caption{Number of Tours per clustering coef.}
\label{fig4}
\end{center}
\end{figure}

As the clustering coefficient gets larger the number of nodes in a tour has to be increased to detect communities.

\section{Results}
\subsection{Results on the Food Network}
In our example we run the algorithm against the Eastern Food Network composed by different ingredients using in the Eastern cuisine. The idea is that the algorithm should be able to find groups of ingredients that are frequently used together. Using $et$ at 0.75 and simulating 150.000 tours of 30 hops the algorithm processed a total of 8308 tours to find clusters with 5 to 10 nodes in less than 5 seconds and these were the top results.

\begin{table}[!hbt]
\center{
\begin{tabular}{|l|lp{7cm}}\hline
\parbox{7cm}{(1909) [orange, vanilla ,almond, orangejuice ,cinnamon, walnut, raisin, honey, butter, cream, milk]} \\ \hline
\parbox{7cm}{(1603):[vanilla, egg, orange, cream, butter, raisin, milk, walnut, cinnamon, orangejuice, almond]} \\ \hline
\parbox{7cm}{(779):[vanilla, butter, egg, cream, milk, almond, raisin, cinnamon, orangejuice, walnut, honey]} \\ \hline
\parbox{7cm}{(704):[vanilla, orangejuice, orange, walnut, raisin, cinnamon, cream, honey, milk, butter, ginger]} \\ \hline
\parbox{7cm}{(534):[cream, vanilla, milk, egg, butter, raisin, walnut, cinnamon, honey, yogurt, ginger]} \\ \hline
\parbox{7cm}{(420):[vanilla, orangejuice, almond, cinnamon, raisin, walnut, butter, honey, cream, milk, ginger]} \\ \hline
\parbox{7cm}{(387):[vanilla, cream, egg, milk, raisin, butter, orangejuice, walnut, cinnamon, honey, yogurt]} \\ \hline
\parbox{7cm}{(187):[vanilla, raisin, walnut, cream, orangejuice, cinnamon, milk, butter, honey, ginger, vinegar]} \\ \hline
\parbox{7cm}{(161):[orange, orangejuice, vanilla, cinnamon, walnut, lemon, raisin, almond, honey, butter, cream]} \\ \hline
\parbox{7cm}{(136):[coriander, garlic, cayenne, bread, lamb, bellpepper, tomato, mint, cucumber, vinegar, beef]} \\ \hline
\end{tabular}
}
\end{table}

The number between parentheses reflects the number of times the same community was detected, so the higher the number the stronger the community.
We can see that the algorithm quickly detects the ingredients for most deserts or breakfast-type preparations. In total the algorithm detected 141 overlapping communities.  The following result looks like a good recipe to try:

\begin{table}[!hbt]
\center{
\begin{tabular}{|l|lp{7cm}}\hline
\parbox{7cm}{(20):[carrot, coriander, chicken, turmeric, ginger, bellpepper, thyme, cumin, chickenbroth, vinegar, cayenne]} \\ \hline
\end{tabular}
}
\end{table}

As a point of comparision we run the modularity optimization algorithm [Blon08] as implemented in Gephi and got the following communities:

[lemon, egg, orange, almond, orangejuice, cream, raisin, cinnamon, honey, butter, milk, vanilla, walnut] 
[coriander, pepper, blackpepper, chicken, thyme, cayenne, cilantro, dill, cumin, bellpepper, chickenbroth, ginger, turmeric, carrot] 
[garlic, parsley, onion, lemonjuice, beef, lamb, tomato, cucumber, bread, oliveoil, mint, vinegar, yogurt, potato] 

As we can see the modularity algorithm does a very good job but it lists all the ingredients that are similar together and is not very helpful to detect smaller groups that go very well together, for example communities of 3 or 4 ingredients. The algorithm presented here would create the following top 10 communities of 3 ingredients:

\begin{table}[!hbt]
\center{
\begin{tabular}{|l|lp{7cm}}\hline
\parbox{7cm}{(1860):[orange, vanilla, almond]} \\ \hline
\parbox{7cm}{(1675):[vanilla, egg, orange]} \\ \hline
\parbox{7cm}{(746):[vanilla, butter, egg]} \\ \hline
\parbox{7cm}{(677):[raisin, orangejuice, orange]} \\ \hline
\parbox{7cm}{(522):[cream, vanilla, milk]} \\ \hline
\parbox{7cm}{(156):[garlic, coriander, cayenne]} \\ \hline
\parbox{7cm}{(95):[coriander, pepper, turmeric]} \\ \hline
\parbox{7cm}{(70):[bread, tomato, cucumber]} \\ \hline
\parbox{7cm}{(43):[bread, lamb, garlic]} \\ \hline
\end{tabular}
}
\end{table}

The graph of the communities found by Gephi looks like this [Figure2]

\begin{figure}[!htb]
\begin{center}
\includegraphics[width=3.5in]{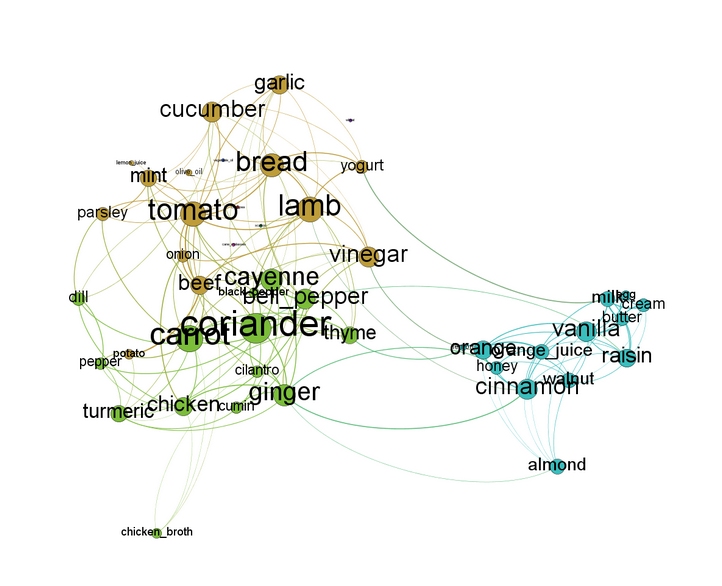}
\caption{Eastern Ingredients.}
\label{fig5}
\end{center}
\end{figure}

As we can see the results help to create new recipes starting with ingredients that go well together frequently. Something interesting is that by allowing overlapping communities we can see that some ingredients are partially in different groups. For example ginger is used for both savory and deserts. The modularity algorithm is forced to choose only one cluster for ginger but in our algorithm we can find it in different communities.

\subsection{Results on Large Social Networks}

We also run the algorithm in a very large dump of a Social Network with a total of about 5 million nodes. The algorithm runs in constant time regardless of the size of the graph as it always simulates a constant number of random walks, the only difference in runtime is due to the time needed to access the adjacency list of each node and that is independant of the clustering algorithm.

Besides the runtime analysis we weree curious to investigate what kind of small communities the algorithm would find in a large Social Network. We run a modularity clustering phase first and then the entropy walker algorithm.

\begin{figure}[!htb]
\begin{center}
\includegraphics[width=4in]{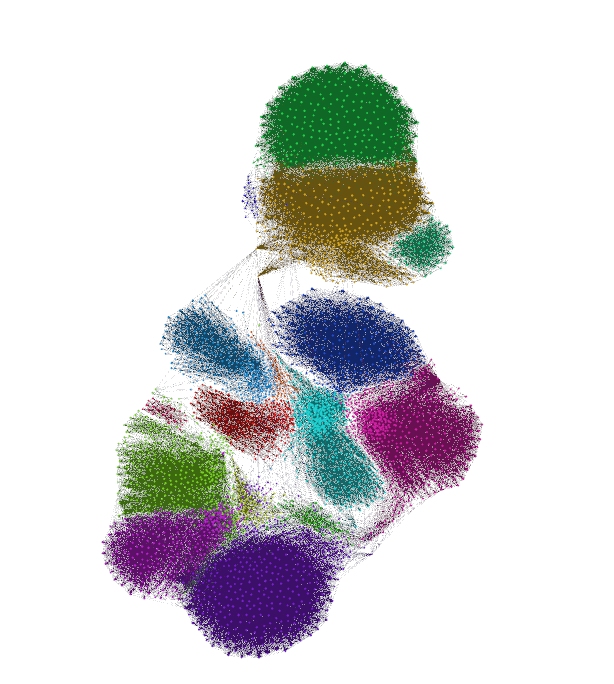}
\caption{Modularity Clustering of the Social Network.}
\label{fig6}
\end{center}
\end{figure}

After running the entropy walker algorithm we found that 100

\begin{figure}[!htb]
\begin{center}
\includegraphics[width=3.5in]{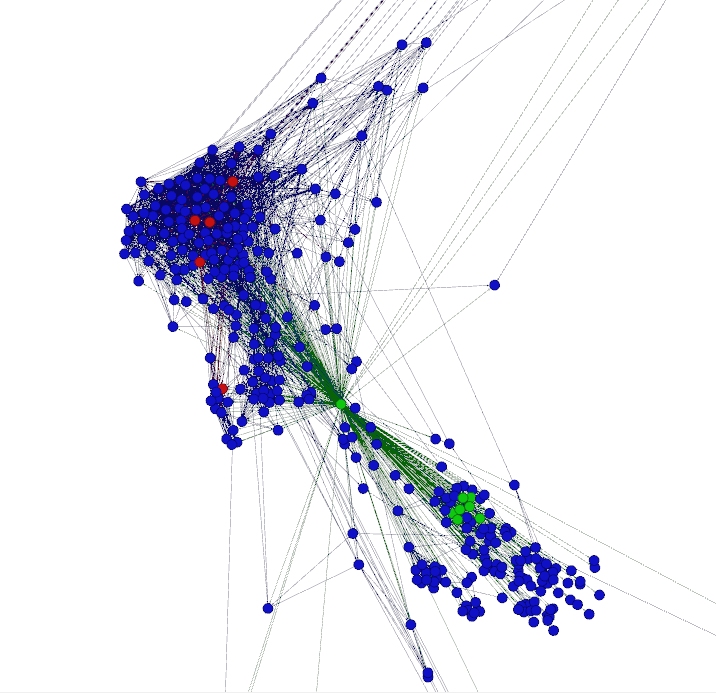}
\caption{An accepted random walk inside a modularity class.}
\label{fig6}
\end{center}
\end{figure}

We see that the entropy walker algorithm finds small dense communities inside the big communities created by the modularity algorithm. 

\begin{figure}[!htb]
\begin{center}
\includegraphics[width=3.5in]{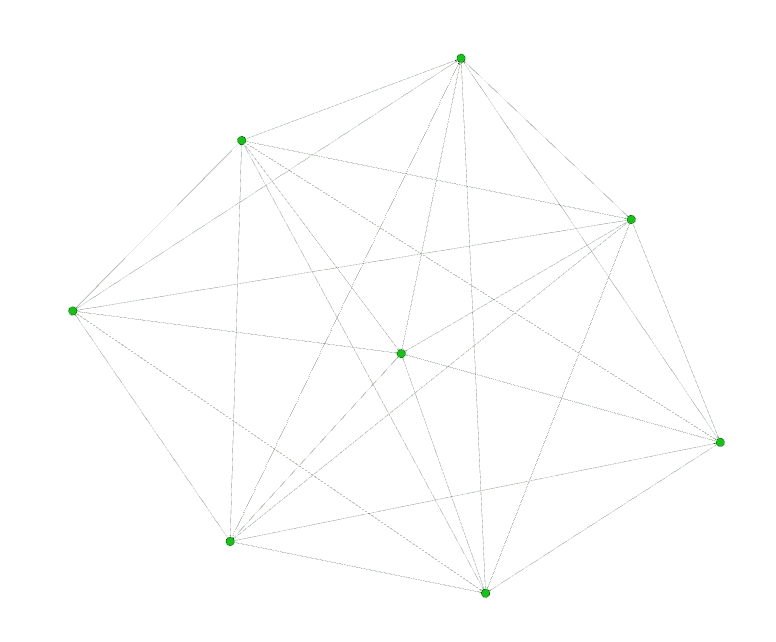}
\caption{Shape of a random walk.}
\label{fig7}
\end{center}
\end{figure}

Figure 4 shows an accepted random walk inside a modularity class. Figure 5 shows the shape of one of the accepted random walk, we can see the community is actually a clique so the algorithm is finding cliques or structures similar to cliques for the parametrized size of components that depend on the length of the random walks.

\section{The Streamming Model}

In a streaming model the graph is constantly updated via the addition and deletion of nodes and edges. In this model the algorithm can be kept running continuously producing "infinite" tours. As the graph is updated communities that were previously detected might disappear and new communities can emerge. An algorithm like the Count-Min Sketch [Mutu05] can be used to keep in memory a list of only the top $n$ communities discovered so far. If a new very tight community forms it will be eventually found by the algorithm several times entering the top $n$ ranking. Besides keeping the top $n$ communities the streaming model can be used to detect communities that pass the entropy filter and the count-min sketch can be used to only list those communities that have repeated a number of times. Several strategies to prune old communities from memory can be used. 

\section{Conclusions}

The entropy walker is a very simple algorithm, the core is just a montecarlo simulation of random walks in a graph. The algorithm uses two very simple tricks to be able to compute communities from these random walks, first it is able to keep or discard a tour by calculating its entropy reasoning that a tour that gets trapped inside a community will visit several times the same nodes resulting in a low-entropy tour. The second trick is the use of LSH and the ability to merge similar tours into a single one to reduce memory consumption and be able to detect the same community even if the nodes have been visited in different order and with different frequencies.

The algorithm can run very quickly consuming very little memory even for massive graphs, it can be kept running continusly in a streamming model where the graph is constantly updated, this setup is perfect for the anlysis of large Social Networks.

\footnotesize
\section{References}

\small{

[Lesk13] Jaewon Yang, Jure Leskovec. (2013) "Overlapping Community Detection at Scale: A Nonnegative Matrix Factorization Approach"
in {\it ACM International Conference on Web Search and Data Mining (WSDM), 2013.}

[B08] Vincent D. Blondel, Jean-Loup Guillaume, Renaud Lambiotte, Etienne Lefebvre. (2008). "Fast unfolding of communities in large networks" 

[vDonn99] Stijn van Dongen. "MCL A cluster algorithm for graphs".in {\it Technical Report INS-R0010, National Research Institute for Mathematics and Computer Science in the Netherlands, Amsterdam, May 2000}

[Mutu05] Graham Cormode and S. Muthukrishnan. (2005) "An Improved Data Stream Summary:
The Count-Min Sketch and its Applications" in {\it Journal of Algorithms archive
Volume 55 Issue 1, April 2005 }

[Mca13] Julian McAuley, Jure Leskovec. (2013) "Discovering Social Circles in Ego Networks" in {\it ACM Trans. Knowl.
Discov. Data 8, 1, Article 4 (February 2014), 28 pages.}

[Bar99] Albert-László Barabási, Réka Albert. (1999) "Emergence of Scaling in Random Networks" in {\it Science 15 October 1999: 
Vol. 286 no. 5439 pp. 509-512}

[Sha48] C. E. Shannon. (1948) "A Mathematical Theory of Communication" in {\it The Bell System Technical Journal vol 27}

}

\end{document}